\documentclass[onecolumn,prd,aps,12pt]{revtex4}%
\usepackage{amsmath}
\usepackage{amsfonts}
\usepackage{amssymb}
\usepackage{graphicx}%
\providecommand{\U}[1]{\protect\rule{.1in}{.1in}}
\begin{document}
\title{The Stark Effect with Minimum Length}
\author{H. L. C. Louzada$^{a}$, H. Belich$^{a}$.}
\affiliation{$^{a}${\small {\ Departamento de F\'{\i}sica e Qu\'{\i}mica, Universidade
Federal do Esp\'{\i}rito Santo, Vit\'{o}ria, ES, 29060-900, Brazil.}}}
\email{belichjr@gmail.com, haofisica@bol.com.br.}
\date{\today}

\begin{abstract}
We will study the splitting in the energy spectrum of the hydrogen atom
subjected to a uniform electric field (Stark effect) with the Heisenberg
algebra deformed leading to the minimum length. We will use the perturbation
theory for cases not degenerate ($n=1$) and degenerate ($n=2$), along with
known results of corrections in these levels caused by the minimum length
applied purely to the hydrogen atom, so that we may find and estimate the
corrections of minimum length applied to the Stark effect.

\end{abstract}
\maketitle

\section{Introduction}

The proposal of noncommutative geometry was developed in 1980 by A. Connes
\cite{33} and it was realized that the non-commutative geometry would be a
scheme to extend the standard model in several ways \cite{necra}. In the 90s
the proposal appears naturally in the context of string theory \cite{34, 44}.
In this way we may obtain an effective theory describing scenarios in string
theory whose in the low energy limit is reduced to a known physical.

Noncommutative geometry also appear, in a condensed matter context, as an
effective theory that describes the electron in a two-dimensional surface
attached to a strong magnetic field. This effective theory describes the
Quantum Hall Effect. The electron would be trapped in the lowest Landau levels
and presents the conductance Hall in $\frac{e^{2}}{\hbar}$ units \cite{56}.

A possible way to explore the implementation of noncommutatives theories is by
the deformation of the Heisenberg algebra. In this paper we study a modified
Heisenberg algebra, by adding certain small corrections to the canonical
commutation relations, it leads, as shown by A. Kempf and contributors
\cite{K1,K2,K3,K4,K5}, to the minimum uncertainty in the position measurement,
$\Delta x_{0},$ called minimum length. The existence of this minimum length
was also suggested by quantum gravity and string theory \cite{DJ,MM,EW}. In
$D$-dimensional case the deformed algebra proposed by Kempf reads:
\begin{align}
\left[  X_{i},P_{j}\right]   &  =i[\delta_{ij}(1+\beta P^{2})+\beta^{^{\prime
}}P_{i}P_{j}],\quad\lbrack P_{i},P_{j}]=0,\label{es1}\\
\left[  X_{i},X_{j}\right]   &  =i\frac{[(2\beta-\beta^{^{\prime}}%
)+(2\beta+\beta^{^{\prime}})\beta P^{2}](P_{i}X_{j}-P_{j}X_{i})}{(1+\beta
P^{2})}, \label{es1a}%
\end{align}

where $\beta$ and $\beta^{^{\prime}}$ are deformation parameters, and we
assume $\beta,\beta^{^{\prime}}>0$. From the uncertainty relation it follows
that the minimum length is $\Delta x_{0}=\sqrt{\beta+\beta^{^{\prime}}}$.

The hydrogen atom is one of the simplest quantum systems that allows
theoretical predictions of high accuracy, and is well-studied experimentally
offering the most precise amount of measures \cite{SG}. There are many papers
where the energy spectrum of the hydrogen atom in the presence minimum length
is calculated \cite{FB,SB,RA}, some of which have divergences in levels $s$
($n=1$) \cite{SB}.

The Stark effect is the splitting of spectral lines of atoms and molecules due
to the action of an external electric field. In our work we will study the
Stark effect with minimum length taking as reference the results of
\cite{MM1}, in which it shows corrections in all energy levels, including $s$,
so that we may find and estimate the corrections of minimum length applied to
the Stark effect.

\section{Hydrogen Atom with Minimum Length}

The eigenvalue problem for the hydrogen atom in $D$ dimensions is described by
the equation%

\begin{equation}
\left(  \frac{P^{2}}{2m}-\frac{e^{2}}{R}\right)  \psi=E\psi, \label{ha}%
\end{equation}
where $e=q$ is the electron charge ($e^{2}=q^{2}/4\pi\epsilon_{0}$ $\left(
\text{SI}\right)  $), the position operators $X_{i}$ and the momentum
operators $P_{i}$ \ satisfies the deformed commutation relations (\ref{es1}),
with $X_{i}=\sqrt{\sum_{i=1}^{D}X_{i}^{2}}$.

We will use the following representation that satisfies the algebra
(\ref{es1}) to first order in $\beta$ and $\beta^{^{\prime}}$%

\begin{align}
X_{i}  &  =x_{i}+\frac{(2\beta-\beta^{^{\prime}})}{4}(x_{i}p^{2}+p^{2}%
x_{i}),\label{es3}\\
P_{i}  &  =p_{i}+\frac{\beta^{^{\prime}}}{2}p_{i}p^{2}, \label{es3a}%
\end{align}
with $p^{2}=\sum_{i=1}^{D}p_{i}^{2}$ and the $x_{i}$ and $p_{i}$ operators
follows the canonical commutation relations $\left[  x_{i},p_{j}\right]
=i\delta_{ij}$. For the undeformed Heisenberg algebra the representation
position may be taken as $x_{i}=x_{i}$ and $p_{i}=i\frac{\partial}{\partial
x^{i}}$.

As it has been shown in \cite{MM1} the analogous Hamiltonian (\ref{ha}) can be
expressed, using the representation (\ref{es3}) and considering only terms of
first order in $\beta$ and $\beta^{^{\prime}},$ as%

\begin{equation}
H=\frac{p^{2}}{2m}+\frac{\beta^{^{\prime}}p^{4}}{2m}-e^{2}\left\{  \frac{1}%
{r}-\frac{(2\beta-\beta^{^{\prime}})}{4}\left[  \frac{1}{r}p^{2}+p^{2}\frac
{1}{r}+\frac{(D-1)}{r^{3}}\right]  \right\}  , \label{hh}%
\end{equation}
and it gives the perturbative correction%

\begin{equation}
\Delta E_{nl}^{(1)}=\frac{e^{2}}{a_{0}^{3}n^{3}}\left[  \frac{(D-1)(2\beta
-\beta^{^{\prime}})}{4\bar{l}(\bar{l}+1)(\bar{l}+1/2)}+\frac{(2\beta
+\beta^{^{\prime}})}{(\bar{l}+1/2)}-\frac{(\beta+\beta^{^{\prime}})}{\bar{n}%
}\right]  ,
\end{equation}
where $a_{0}$ is the Bohr radius, $\bar{n}=n+(D-3)/2$, $\bar{l}=l+(D-3)/2$,
with the principal quantum number represented by $n$ and the orbital quantum
number by $l$. This expression is singular in $D=3$ and $l=0$, and, in the
same paper, it was shown that the Hamiltonian (\ref{hh}) can also be expressed as%

\begin{equation}
H=\frac{p^{2}}{2m}+\frac{\beta^{^{\prime}}p^{4}}{2m}-e^{2}\left[  \frac
{1}{\sqrt{r^{2}+b^{2}}}-\frac{(2\beta-\beta^{^{\prime}})}{4}\left(  \frac
{1}{r}p^{2}+p^{2}\frac{1}{r}\right)  \right]  , \label{es6}%
\end{equation}
with $b=\sqrt{2\beta-\beta^{^{\prime}}}$. So the correction for $n=1$ is given by%

\begin{equation}
\Delta E_{1s}^{(1)}=\left\langle 1,0,0|V|1,0,0\right\rangle ,
\end{equation}
where%

\begin{equation}
V=\frac{\beta^{^{\prime}}p^{4}}{2m}-e^{2}\left[  \frac{1}{\sqrt{r^{2}+b^{2}}%
}-\frac{1}{r}-\frac{(2\beta-\beta^{^{\prime}})}{4}\left(  \frac{1}{r}%
p^{2}+p^{2}\frac{1}{r}\right)  \right]  , \label{es7}%
\end{equation}
it results%

\begin{equation}
\Delta E_{1s}^{(1)}=\frac{e^{2}}{a_{0}^{3}}\left\{  3\beta+\beta^{^{\prime}%
}-(2\beta-\beta^{^{\prime}})\left[  ln\frac{(2\beta-\beta^{^{\prime}})}%
{a_{0}^{2}}+2\gamma+1\right]  \right\}  , \label{es8}%
\end{equation}
where $\gamma=0,57721$ is the Euler constant. The correction to the level $2s$
is expressed by%

\begin{equation}
\label{es9}\Delta E_{2s}^{(1)}=\frac{e^{2}}{8a_{0}^{3}}\left\{  \frac
{(7\beta+3\beta^{^{\prime}})}{2}-(2\beta-\beta^{^{\prime}})\left[
ln\frac{(2\beta-\beta^{^{\prime}})}{4a_{0}^{2}}+2\gamma+\frac{5}{2}\right]
\right\}  .
\end{equation}

This results will be of fundamental importance for our work.

\section{The Stark Effect}

\subsection{The Ordinary Stark Effect}

The splitting of the energy levels of the atoms of an occupation electron (the
hydrogen atom or atoms of the "hydrogen type" with one valence electron
outside a spherically symmetric shell) subjected to an electric field uniform
is said Stark effect \cite{SK}.

The Hamiltonian of the hydrogen atom subjected to a uniform electric field in
the positive direction of the $z$ axis is given by
\begin{equation}
H=\frac{p^{2}}{2m}-\frac{e^{2}}{r}-e|\vec{E}|z,
\end{equation}
which can be divided into two parts, $H=H_{0}+V$ where
\begin{equation}
H_{0}=\frac{p^{2}}{2m}-\frac{e^{2}}{r},
\end{equation}
with
\begin{equation}
V=-e|\vec{E}|z.
\end{equation}
We can treat $V$ as a perturbation. Assuming the eigenstates and $H_{0}$
energy spectrum are known and disregarding the degree of freedom of spin, our
analysis is divided into corrections for non-degenerate energy levels (only
$n=1$) and degenerates ($n\neq1$).

\subsubsection{The Quadratic Stark Effect}

For $n=1$ the energy shift is given by
\begin{equation}
\Delta_{1}=-e|\vec{E}|\left\langle 1,0,0|z|1,0,0\right\rangle +e^{2}|\vec
{E}|^{2}\sum_{n\neq1}\sum_{l,m}\frac{|\left\langle n,l,m|z|1,0,0\right\rangle
|^{2}}{E_{1}-E_{n}}+...\quad,
\end{equation}
we note that the operator $z$ is odd parity while the ground state has
well-defined parity (even) so $\left\langle 1,0,0|z|1,0,0\right\rangle =0$,
and energy shift is quadratic in $|\vec{E}|$, therefore this type of
displacement is said quadratic Stark effect.

~~~~ The polarizability $\alpha$ of an atom is defined in terms of the energy
shift of the atomic state as follows%

\begin{equation}
\Delta_{1}=-\frac{1}{2}\alpha|\vec{E}|^{2},
\end{equation}
in our case
\begin{equation}
\alpha=-2e^{2}\sum_{n\neq1}\sum_{l,m}\frac{|\left\langle
n,l,m|z|1,0,0\right\rangle |^{2}}{E_{1}-E_{n}}.
\end{equation}

We remark that
\begin{align}
\sum_{n\neq1}\sum_{l,m}|\left\langle n,l,m|z|1,0,0\right\rangle |^{2}  &
=\sum_{n,l,m}|\left\langle n,l,m|z|1,0,0\right\rangle |^{2}\nonumber\\
&  =\left\langle 1,0,0|z^{2}|1,0,0\right\rangle =a_{0}^{2}.
\end{align}
However, the denominator is not constant, we can get an upper limit for the
polarizability taking into account that
\begin{equation}
E_{n}=-\frac{e^{2}}{2a_{0}n^{2}},
\end{equation}
so
\begin{equation}
-E_{1}+E_{n}=\frac{e^{2}}{2a_{0}}\left(  1-\frac{1}{n^{2}}\right)  \geq
-E_{1}+E_{2}=\frac{3e^{2}}{8a_{0}},
\end{equation}
then
\begin{equation}
\frac{1}{-E_{1}+E_{n}}\leq\frac{8a_{0}}{e^{2}},
\end{equation}
and therefore%
\begin{equation}
\alpha<\frac{16a_{0}^{3}}{3},
\end{equation}

that is
\begin{equation}
\Delta_{1}>-\frac{8}{3}a_{0}^{3}|\vec{E}|^{2}. \label{es32}%
\end{equation}

\subsubsection{The Linear Stark Effect}

~~~~ For $n\neq1$ occurs degeneracy due to possible values that can be assumed
by $l$, i.e., $0,1,...,n-1$. We focus on the $n=2$ level, then there is a
state with $l=0$ ($|2,0,0>$) said $2s$ state, and three states with $l=1$
($|2,1,0>$, $|2,1,1>$ and $|2,1,-1>$) said $2p$ states, all with the same
energy $E_{2}=-e^{2}/8a_{0}$. The state $|2,0,0>$ is even parity, while three
states $|2,1,m>$ ($m=0,\pm1$) are odd parity, therefore the matrix element
$\left\langle 2,0,0|V|2,0,0\right\rangle $ and the nine matrix elements
$\left\langle 2,1,m^{^{\prime}}|V|2,1,m\right\rangle $ are zero, the only
elements that can be nonzero are of the form $\left\langle 2,1,m^{^{\prime}%
}|V|2,0,0\right\rangle $. Explicitly we have%

\begin{equation}
\left\langle 2,1,0|V|2,0,0\right\rangle =-3ea_{0}|\vec{E}|,
\end{equation}
and
\begin{equation}
\left\langle 2,1,\pm1|V|2,0,0\right\rangle =0.
\end{equation}

The matrix $V$ represents the $n=2$ level then assumes the following form (the
basis vectors are arranged in the following order: $|2,0,0>$, $|2,1,0>$,
$|2,1,1>$ and $|2,1,-1>$ )%

\begin{equation}
V=\left[
\begin{array}
[c]{cccc}%
0 & -3ea_{0}|\vec{E}| & 0 & 0\\
-3ea_{0}|\vec{E}| & 0 & 0 & 0\\
0 & 0 & 0 & 0\\
0 & 0 & 0 & 0
\end{array}
\right]  .
\end{equation}

Proceeding with the diagonalization we get
\begin{equation}
V|\pm>=\mp3ea_{0}|\vec{E}||\pm>,
\end{equation}
with
\begin{equation}
|\pm>=\frac{1}{\sqrt{2}}(|2,0,0>\pm|2,1,0>), \label{es23a}%
\end{equation}
and
\begin{equation}
V|2,1,1>=V|2,1,-1>=0.
\end{equation}
We see then that the degeneracy of the $n=2$ level is partially removed and
energy shift is linear in $\vec{E}$ and, for this reason, it is said linear
Stark effect.

\subsection{The Stark Effect of Minimum Length}

\subsubsection{The Quadratic Stark Effect}

Using (\ref{es3}) and (\ref{es6}), the Hamiltonian of the hydrogen atom
subjected to a uniform electric field in the positive direction of the $z$
axis is%

\begin{equation}
H=\frac{p^{2}}{2m}+\frac{\beta^{^{\prime}}p^{4}}{2m}-e^{2}\left[  \frac
{1}{\sqrt{r^{2}+b^{2}}}-\frac{(2\beta-\beta^{^{\prime}})}{4}\left(  \frac
{1}{r}p^{2}+p^{2}\frac{1}{r}\right)  \right]  -e|\vec{E}|\left[
z+\frac{(2\beta-\beta^{^{\prime}})}{4}(zp^{2}+p^{2}z)\right]  .
\end{equation}
The total energy splitting for $n=1,$ for first order in $\beta$ and
$\beta^{^{\prime}},$\ is given by
\begin{equation}
\Delta E_{1s}^{(T)}=\Delta E_{1s}^{(1)}+\Delta_{1}^{ML},
\end{equation}
where $\Delta E_{1s}^{(1)}$ is given by ($\ref{es8}$) and
\begin{equation}
\Delta_{1}^{ML}=-e|\vec{E}|\left\langle 1,0,0|Z|1,0,0\right\rangle +e^{2}%
|\vec{E}|^{2}\sum_{n\neq1}\sum_{l,m}\frac{|\left\langle
n,l,m|Z|1,0,0\right\rangle |^{2}}{E_{1}-E_{n}}+...\quad,
\end{equation}
with
\begin{equation}
Z=z+\frac{(2\beta-\beta^{^{\prime}})}{4}(zp^{2}+p^{2}z).
\end{equation}

The term $\left\langle 1,0,0|Z|1,0,0\right\rangle $ will also be null, since
$zp^{2}$ and $p^{2}z$ are also odd parity operators. Similarly
\begin{equation}
\Delta_{1}^{ML}=-\frac{1}{2}\alpha|\vec{E}|^{2},
\end{equation}
we know that
\begin{equation}
\sum_{n\neq1}\sum_{l,m}|\left\langle n,l,m|Z|1,0,0\right\rangle |^{2}%
=\left\langle 1,0,0|Z^{2}|1,0,0\right\rangle ,
\end{equation}
and after some calculations we obtain that
\begin{equation}
\left\langle 1,0,0|Z^{2}|1,0,0\right\rangle =a_{0}^{2}+\frac{(2\beta
-\beta^{^{\prime}})}{2},
\end{equation}
then
\begin{equation}
\alpha<\frac{16}{3}a_{0}\left[  a_{0}^{2}+\frac{(2\beta-\beta^{^{\prime}})}%
{2}\right]  ,
\end{equation}
i. e.
\begin{equation}
\Delta_{1}^{ML}>-\frac{8}{3}a_{0}\left[  a_{0}^{2}+\frac{(2\beta
-\beta^{^{\prime}})}{2}\right]  |\vec{E}|^{2}. \label{es23}%
\end{equation}

\subsubsection{The Linear Stark Effect}

Here the total Hamiltonian is%

\begin{equation}
H=H_{0}+V_{ML}+V_{ML}^{S},
\end{equation}
where $H_{0}$ is the Hamiltonian of hydrogen atom, $V_{ML}$ is the
perturbation caused only by the minimum length and $V_{ML}^{S}$ is the
perturbation caused by the minimum length applied to the Stark effect. For the
matrix elements of the form $\left\langle n,0,0|V_{ML}|n^{^{\prime}%
},0,0\right\rangle $ the explicit formula to the perturbation is%

\begin{equation}
V_{ML}=\frac{\beta^{^{\prime}}p^{4}}{2m}-e^{2}\left[  \frac{1}{\sqrt
{r^{2}+b^{2}}}-\frac{1}{r}-\frac{(2\beta-\beta^{^{\prime}})}{4}\left(
\frac{1}{r}p^{2}+p^{2}\frac{1}{r}\right)  \right]  , \label{es34}%
\end{equation}
whereas for elements of the form $\left\langle n,l,m|V_{ML}|n^{^{\prime}%
},l^{^{\prime}},m^{^{\prime}}\right\rangle $ it is given by \cite{MM2}
\begin{equation}
V_{ML}=\frac{\beta^{^{\prime}}p^{4}}{2m}+\frac{(2\beta-\beta^{^{\prime}})}%
{4}e^{2}\left(  \frac{1}{r}p^{2}+p^{2}\frac{1}{r}+\frac{2}{r^{3}}\right)  .
\label{es35}%
\end{equation}
Besides that we have%

\begin{equation}
V_{ML}^{S}=-e|\vec{E}|\left[  z+\frac{(2\beta-\beta^{^{\prime}})}{4}%
(zp^{2}+p^{2}z)\right]  ,
\end{equation}
which is also odd parity, therefore the $\left\langle 2,0,0|V_{ML}%
^{S}|2,0,0\right\rangle $ and the nine matrix elements $\left\langle
2,1,m^{^{\prime}}|V_{ML}^{S}|2,1,m\right\rangle $ will also be null. After
some calculations we obtain that%

\begin{equation}
\left\langle 2,1,0|V_{ML}^{S}|2,0,0\right\rangle =-e|\vec{E}|\left[
3a_{0}-\frac{(2\beta-\beta^{^{\prime}})}{8a_{0}}\right]  ,
\end{equation}
with $|\pm>$ given by (\ref{es23a}) and
\begin{equation}
\left\langle 2,1,\pm1|V_{ML}^{S}|2,0,0\right\rangle =0.
\end{equation}

Similarly to the case with no minimum length, diagonalizing the matrix
$V_{ML}^{S}$ we get%

\begin{equation}
V_{ML}^{S}|\pm>=\mp e|\vec{E}|\left[  3a_{0}-\frac{(2\beta-\beta^{^{\prime}}%
)}{8a_{0}}\right]  |\pm>, \label{es39}%
\end{equation}
and
\begin{equation}
V_{ML}^{S}|2,1,1>=V_{ML}^{S}|2,1,-1>=0.
\end{equation}

The elements of the matrix $V_{ML}$ are given explicitly by \cite{MM2}
\begin{equation}
\left\langle n^{^{\prime}},l^{^{\prime}},m^{^{\prime}}|V_{ML}%
|n,l,m\right\rangle =\delta_{ll^{^{\prime}}}\delta_{mm^{^{\prime}}}\left(
2m\beta^{^{\prime}}E_{n}^{2}\delta_{nn^{^{\prime}}}-\left\langle n^{^{\prime}%
},l,m|V_{ML}^{^{\prime}}|n,l,m\right\rangle \right)  ,
\end{equation}
with $V_{ML}^{^{\prime}}$ given by
\begin{equation}
V_{ML}^{^{\prime}}=-e^{2}\left[  \frac{1}{\sqrt{r^{2}+b^{2}}}-\frac{1}%
{r}-\frac{(2\beta-\beta^{^{\prime}})}{4}\left(  \frac{1}{r}p^{2}+p^{2}\frac
{1}{r}\right)  \right]  ,
\end{equation}
for elements of type $\left\langle n^{^{\prime}},0,0|V_{ML}^{^{\prime}%
}|n,0,0\right\rangle $, and
\begin{equation}
V_{ML}^{^{\prime}}=\frac{(2\beta-\beta^{^{\prime}})}{4}e^{2}\left(  \frac
{1}{r}p^{2}+p^{2}\frac{1}{r}+\frac{2}{r^{3}}\right)  ,
\end{equation}
for elements of type $\left\langle n^{^{\prime}},l^{^{\prime}},m^{^{\prime}%
}|V_{ML}^{^{\prime}}|n,l,m\right\rangle $. We then see that, for
$n=n^{^{\prime}}=2$, the matrix $V_{ML}$ is diagonal, i.e., it is diagonal in
the basis $\left\{  |2,0,0>,|2,1,0>,|2,1,1>,|2,1,-1>\right\}  $, but not in
the basis $\left\{  |+>,|->,|2,1,1>,|2,1,-1>\right\}  $, in which it gives
effect linear Stark effect with minimum length. Therefore, the perturbation
$V_{ML}$, purely caused by the minimum length, and the perturbation
$V_{ML}^{S}$, caused by the minimum length applied to the Stark effect can not
be simultaneously measured.

\section{Estimated correction}

We can rewrite (\ref{es23}) using the parameters%

\begin{equation}
\Delta x_{min}=\sqrt{\beta+\beta^{^{\prime}}},
\end{equation}
and
\begin{equation}
\eta=\frac{\beta}{(\beta+\beta^{^{\prime}})},
\end{equation}
instead of $\beta$ and $\beta^{^{\prime}}$, as done in \cite{SB,MM1,MM3,MM4},
with $1/3\leq\eta\leq1$. Once this is done we get
\begin{equation}
\Delta_{1}^{ML}>-\frac{8}{3}a_{0}\left[  a_{0}^{2}+\frac{\Delta x_{min}%
^{2}(3\eta-1)}{2}\right]  |\vec{E}|^{2},
\end{equation}
or
\begin{equation}
\Delta_{1}^{ML}>-\frac{8}{3}a_{0}^{3}|\vec{E}|^{2}+\varsigma(\eta),
\end{equation}
where
\begin{equation}
\varsigma(\eta)=-\frac{4}{3}a_{0}|\vec{E}|^{2}\Delta x_{min}^{2}(3\eta-1)
\end{equation}
is the correction caused by the minimum length applied to the quadratic Stark
effect. We clearly see that for $\eta=1/3$ the correction is null and to
$\eta=1$ the correction is maximum (in magnitude). Taking into account the
value $\Delta x_{min}(\eta=1)=2,86\cdot10^{-17}m$ obtained in \cite{MM1}
(Attributing the difference between the experimental and theoretical values of
the Lamb shift for the $1s$ level of the hydrogen atom entirely to the minimum
length), and the electric field $|\vec{E}|=10^{7}V/m$ (as the usual fields
used in Stark effect experiments are of the order of $10^{6}-10^{7}V/m$
\cite{WG}), we obtain%

\begin{equation}
\varsigma=-1,283\cdot10^{-39}J. \label{es48}%
\end{equation}
This value is $10$ orders of magnitude smaller than the difference between the
experimental and theoretical values of the Lamb shift for the $1s$ level of
the hydrogen atom, $\Delta L_{1s}=7,024\cdot10^{-29}J$, and $12$ orders of
magnitude smaller than the correction of the ordinary quadratic Stark effect
(for the same value of the electric field) $\Delta_{1}=-4,390\cdot10^{-27}J$.

The same can be done for (\ref{es39}) setting%
\begin{equation}
\chi(\eta)\leq-\frac{e}{8a_{0}}|\vec{E}|^{2}\Delta x_{min}^{2}(3\eta-1)
\end{equation}
as the correction caused by the minimum length applied to the linear Stark
effect. Then we obtain%

\begin{equation}
\chi\leq-6,193\cdot10^{-36}J, \label{es49}%
\end{equation}
which is $7$ orders of magnitude less than $\Delta L_{1s}$ and $14$ orders of
magnitude smaller than the correction of the ordinary linear Stark effect,
$\Delta_{1}=-2,542\cdot10^{-22}J$.

\section{Concluding remarks}

We have studied the quadratic and linear Stark effect in the hydrogen atom
taking into account the minimum length. For this goal we have used
perturbation methods, along with known results of corrections in the energy
levels of the hydrogen atom caused by the minimum length \cite{MM1}. Estimates
values obtained from the minimum length applied to the quadratic and linear
Stark effect, (\ref{es48}) and (\ref{es49}), are many orders of magnitude
smaller than the corrections of the ordinary Stark effect. They are also much
smaller than the difference between the experimental and theoretical values of
the Lamb shift for the $1s$ level of the hydrogen atom (which is where you get
the best estimate for the minimum length \cite{SB}).

We observe also that the estimated minimum length correction applied to the
linear Stark effect, which occurs for $n=2$, we use $\Delta L_{1s}%
=7,024\cdot10^{-29}J$, instead of $\Delta L_{2s-2p}=7,951\cdot10^{-30}J$,
which would be strictly correct value. This was done intentionally to
illustrate that even in the \textquotedblleft best
hypothesis\textquotedblright\ because $\Delta L_{1s}>\Delta L_{2s-2p}$, the
obtained correction is very small, clarifying the discussion in the section
(III.B.2) on the corrections caused by the pure minimum length and the minimum
length applied to the linear Stark effect can not be simultaneously measured,
the latter can simply be discarded for practical purposes.

It is interesting to mention that our correction estimate is pretty
simplistic, since the hypothesis attribute the difference between the
experimental and theoretical values of the Lamb shift of the hydrogen atom
entirely to the minimum length is naive compared to the recent discussions on
the proton radius \cite{AA} and proposed models to provide an explanation that
does not conflict with the experimental data \cite{CE,BB}.

\textbf{Ackowledgements}

The Authors thank CNPq and CAPES (of Brazil) for financial support.


\begin{thebibliography}{99}                                                                                               %


\bibitem {33}A. Connes, C* algebras and differential geometry, Compt. Rend,
Acad. Sci. (Ser. I Math), A290:599-609 (1980).

\bibitem {necra}Michael R. Douglas, Nikita A. Nekrasov,
Rev.Mod.Phys.73:977-1029, (2001).

\bibitem {34}A. Connes, Michael R. Douglas and Albert Schwarz, JHEP
\textbf{02}:003 (1998).

\bibitem {44}Michael R. Douglas, Christopher M. Hull, JHEP \textbf{02}:008
(1998). .

\bibitem {56}Quantum Field Theory in Condensed Matter Physics, Naoto Nagaosa
and S. Heusler (1999).

\bibitem {K1}A. Kempf, J. Math Phys. \textbf{35}, 4483 (1994).

\bibitem {K2}A. Kempf, G. Mangano and R. B. Mann, Phys. Rev. D \textbf{52},
1108 (1995).

\bibitem {K3}H. Hinrichsen and A. Kemph, J. Math Phys. \textbf{37}, 2121 (1996).

\bibitem {K4}A. Kempf, J. Math Phys. \textbf{38}, 1347 (1997).

\bibitem {K5}A. Kempf, J. Phys. A \textbf{30}, 2093 (1997).

\bibitem {DJ}D. J. Gross and P.F. Mende, Nucl. Phys. B \textbf{303}, 407 (1988).

\bibitem {MM}M. Maggiore, Phys. Lett. B \textbf{304}, 65 (1993).

\bibitem {EW}E. Witten, Phys. Today \textbf{49}, 24 (1996).

\bibitem {SG}S. G. Karshenboim, Phys. Rep. \textbf{422}, 1 (2005).

\bibitem {FB}F. Brau, J. Phys. A \textbf{32}, 7691 (1999).

\bibitem {SB}S. Benczik, L. N. Chang, D. Minic and T. Takeuchi, Phys. Rev. A
\textbf{72}, 012104 (2005).

\bibitem {RA}R. Akhoury, Y.-P. Yao, Phys. Lett. B \textbf{572}, 37 (2003).

\bibitem {MM1}M. M. Stetsko and V. M. Tkachuk, Phys. Rev. A \textbf{74},
012101 (2006).

\bibitem {SK}J. J. Sakurai and S. Samuel, Modern Quantum Mechanics
(Addison-Wesley Publishing Company, New York, 1994).

\bibitem {MM2}M. M. Stetsko and V. M. Tkachuk, Phys. Lett. A \textbf{372},
5126-5130 (2008).

\bibitem {MM3}M. M. Stetsko, Phys. Rev. A \textbf{74}, 062105 (2006).

\bibitem {MM4}M. M. Stetsko and V. M. Tkachuk, Phys. Rev. A \textbf{76},
012707 (2007).

\bibitem {WG}W. Greiner, Quantum Mechanics An Introduction (Fourth Edition,
Springer, New York, 2001).

\bibitem {AA}A. Antognini et al., Science \textbf{39}, 417-420 (2013).

\bibitem {CE}C. E. Carison and B. C. Rislow, Phys. Rev. D \textbf{86}, 035013 (2012).

\bibitem {BB}B. Batell, D.McKeen and M. Pospelov, Phys. Lett. \textbf{107},
0118013 (2011).
\end{thebibliography}
\end{document}